\documentclass[pra,twocolumn,showpacs,superscriptaddress,floatfix, nofootinbib]{revtex4-2}
\usepackage[caption=false]{subfig}
\usepackage{amsmath,graphicx}
\usepackage{amsmath,amssymb,mathrsfs,esint}
\usepackage{multirow}
\usepackage{algorithm}
\usepackage[noend]{algpseudocode}
\usepackage{comment}
\usepackage{tabularx}
\usepackage{bm}
\usepackage[normalem]{ulem}
\usepackage[bookmarks=true,
   colorlinks=true,
   linkcolor=blue,
   urlcolor=blue,
   citecolor=blue,
   bookmarks=true,
   hyperindex=true
]{hyperref}

\usepackage{tikz, adjustbox, listings}
\usetikzlibrary{quantikz2}

\definecolor{ForestGreen}{RGB}{34,139,34}

\newcommand{\GREEN}[1]{{\color{black}{#1}}}
\newcommand{\BLUE}[1]{{\color{black}{#1}}}

\begin{document}

\title{Supervised binary classification of small-scale digit images \BLUE{and weighted graphs} \\ with a trapped-ion quantum processor}

\author{Ilia V. Zalivako}
\affiliation{P.N. Lebedev Physical Institute of the Russian Academy of Sciences, Moscow 119991, Russia}
\affiliation{Russian Quantum Center, Skolkovo, Moscow 121205, Russia}

\author{Alexander I.~Gircha}
\affiliation{P.N. Lebedev Physical Institute of the Russian Academy of Sciences, Moscow 119991, Russia}
\affiliation{Russian Quantum Center, Skolkovo, Moscow 121205, Russia}

\author{Evgeniy O. Kiktenko}
\affiliation{P.N. Lebedev Physical Institute of the Russian Academy of Sciences, Moscow 119991, Russia}
\affiliation{Russian Quantum Center, Skolkovo, Moscow 121205, Russia}

\author{Anastasiia S. Nikolaeva}
\affiliation{P.N. Lebedev Physical Institute of the Russian Academy of Sciences, Moscow 119991, Russia}
\affiliation{Russian Quantum Center, Skolkovo, Moscow 121205, Russia}

\author{Denis A. Drozhzhin}
\affiliation{P.N. Lebedev Physical Institute of the Russian Academy of Sciences, Moscow 119991, Russia}
\affiliation{Russian Quantum Center, Skolkovo, Moscow 121205, Russia}

\author{Alexander S. Borisenko}
\affiliation{P.N. Lebedev Physical Institute of the Russian Academy of Sciences, Moscow 119991, Russia}
\affiliation{Russian Quantum Center, Skolkovo, Moscow 121205, Russia}

\author{Andrei E. Korolkov}
\affiliation{P.N. Lebedev Physical Institute of the Russian Academy of Sciences, Moscow 119991, Russia}
\affiliation{Russian Quantum Center, Skolkovo, Moscow 121205, Russia}

\author{Nikita V. Semenin}
\affiliation{P.N. Lebedev Physical Institute of the Russian Academy of Sciences, Moscow 119991, Russia}
\affiliation{Russian Quantum Center, Skolkovo, Moscow 121205, Russia}

\author{Kristina P. Galstyan}
\affiliation{P.N. Lebedev Physical Institute of the Russian Academy of Sciences, Moscow 119991, Russia}
\affiliation{Russian Quantum Center, Skolkovo, Moscow 121205, Russia}

\author{Pavel A. Kamenskikh}
\affiliation{P.N. Lebedev Physical Institute of the Russian Academy of Sciences, Moscow 119991, Russia}
\affiliation{Russian Quantum Center, Skolkovo, Moscow 121205, Russia}

\author{Vasilii N. Smirnov}
\affiliation{P.N. Lebedev Physical Institute of the Russian Academy of Sciences, Moscow 119991, Russia}
\affiliation{Russian Quantum Center, Skolkovo, Moscow 121205, Russia}

\author{Mikhail A. Aksenov}
\affiliation{Russian Quantum Center, Skolkovo, Moscow 121205, Russia}

\author{Pavel L. Sidorov}
\affiliation{P.N. Lebedev Physical Institute of the Russian Academy of Sciences, Moscow 119991, Russia}
\affiliation{Russian Quantum Center, Skolkovo, Moscow 121205, Russia}

\author{Ksenia Yu. Khabarova}
\affiliation{P.N. Lebedev Physical Institute of the Russian Academy of Sciences, Moscow 119991, Russia}
\affiliation{Russian Quantum Center, Skolkovo, Moscow 121205, Russia}

\author{Aleksey K. Fedorov}
\affiliation{P.N. Lebedev Physical Institute of the Russian Academy of Sciences, Moscow 119991, Russia}
\affiliation{Russian Quantum Center, Skolkovo, Moscow 121205, Russia}

\author{Nikolay N. Kolachevsky}
\affiliation{P.N. Lebedev Physical Institute of the Russian Academy of Sciences, Moscow 119991, Russia}
\affiliation{Russian Quantum Center, Skolkovo, Moscow 121205, Russia}

\author{Ilya A. Semerikov}
\affiliation{P.N. Lebedev Physical Institute of the Russian Academy of Sciences, Moscow 119991, Russia}
\affiliation{Russian Quantum Center, Skolkovo, Moscow 121205, Russia}

\begin{abstract}
Here we present the results of benchmarking a quantum processor based on trapped $^{171}$Yb$^{+}$ ions by performing basic quantum machine learning algorithms. 
\BLUE{Using a quantum-enhanced support vector machine algorithm with up to five qubits we perform a supervised binary classification on two types of datasets: small binary digit images and weighted graphs with a ring topology.
For the first dataset, images are intentionally selected so that they could be classified with 100\% accuracy. 
This allows us to specifically examine different types of quantum encodings of the digit dataset and study the impact of experimental noise.
In the second dataset, graphs are divided into two categories based on the spectral structure of their Ising Hamiltonian models, which is related to the NP-hard problem. 
For this problem we consider an embedding of an exponentially large Hamiltonian spectrum into an entangled state of a linear number of qubits.
For both problems, we study various levels of circuit optimization and found that, for all experiments conducted, we achieve classifiers with 100\% accuracy on both training and testing datasets. 
This demonstrates that the quantum processor has the ability to correctly solve the basic classification task under consideration.} As we expect, with the increase in the capabilities of quantum processors, they can be utilized for solving machine learning tasks.
\end{abstract}

\maketitle

\section{Introduction}

Recent progress in developing quantum computing devices has shown their potential to solve computational problems at the threshold of capabilities of computing devices based on classical principles~\cite{Martinis2019,Pan2020,Pan2021,Lukin2022}. 
Various physical platforms for quantum computing, 
such as superconducting circuits~\cite{Martinis2019,Pan2021-4}, semiconductor quantum dots~\cite{Vandersypen2022,Morello2022,Tarucha2022}, photonic systems~\cite{Pan2020,Lavoie2022}, 
neutral atoms~\cite{Lukin2021,Browaeys2021,Browaeys2020-2,Saffman2022}, 
and trapped ions~\cite{Monroe2017,Blatt2012,Blatt2018},
are currently under development. 
Although such quantum devices are used to solve certain classes of computational problems~\cite{Saffman2022,Blatt2018,Lukin2021,Browaeys2021,Browaeys2020-2,Monroe2017,Blatt2012},
demonstration of a sizable computational advantage in solving practical problems remains a challenge~\cite{Fedorov2022}.
This quest for practical quantum computational advantage poses interesting problems.
On the one hand, one needs to increase the computational capabilities of quantum computers, which requires not only scaling to the significant number of qubits but also improving the quality of quantum operations (i.e., quantum gate fidelities). 
From this point of view, trapped-ion-based quantum processors demonstrate the highest quantum volume of $2^{21}$ in experiments by Quantinuum~\cite{Quantinuum2024}.
In addition, trapped-ion-based quantum devices have been used to demonstrate error correction~\cite{Wineland2004,Blatt2011,Blatt2020,Monroe2021,Blatt2021-2,Postler2022}, 
e.g., a fault-tolerant entanglement between two logical qubits~\cite{Postler2022,Ryan-Anderson2022} and quantum algorithms with logical qubits~\cite{Yamamoto2023} have been realized. 
Therefore, such systems seem to be promising for running quantum algorithms~\cite{Monroe2016,Blatt2016,Fedorov2022}.

On the other hand, a problem to solve on a quantum processor should be chosen carefully.
Candidates include integer factorization~\cite{Shor1994} and simulating complex quantum systems~\cite{Lloyd1996}.
However, achieving quantum advantage in these directions seems to require computational resources that are far beyond the capabilities of the upcoming generation of quantum devices~\cite{Fedorov2022}.
One of the directions, which is under exploration in the context of near-term applications, is the field of quantum machine learning~\cite{Biamonte2017}. 
Various approaches have been proposed.
In particular, quantum convolutional neural networks~\cite{Henderson2020-2,Lukin2019-3,Bokhan2022}, 
generative adversarial networks~\cite{Lloyd2018,Dallaire-Demers2018,Woerner2019},
kernel methods~\cite{Chatterjee2017,Havlek2018SupervisedLW,Schuld2021qmodels}, and other approaches have been studied. 
Recent experimental works in the field of quantum machine learning~\cite{Gambetta2017-3} include classifiers for handwritten digits datasets~\cite{Benedetti2018,Kerenidis2020,Bokhan2022},
analyzing NMR readings~\cite{Demler2020,Demler2021}, 
classification of lung cancer patients~\cite{Jain2020}, 
classifying and ranking DNA to RNA transcription factors~\cite{Lidar2018-2}, 
satellite imagery analysis~\cite{Henderson2020}, 
generative chemistry~\cite{Fedorov2023}
weather forecasting~\cite{Rigetti2021}, 
and many others (for a review, see Refs.~\cite{Biamonte2017,Perdomo-Ortiz2018,Dunjko2018}). 
Therefore, benchmarking quantum processors under development using quantum machine learning tasks seems to be useful as a step towards demonstrating quantum computational advantage for practically-relevant problems.

\begin{figure}
    \centering
    \includegraphics[width=1\linewidth]{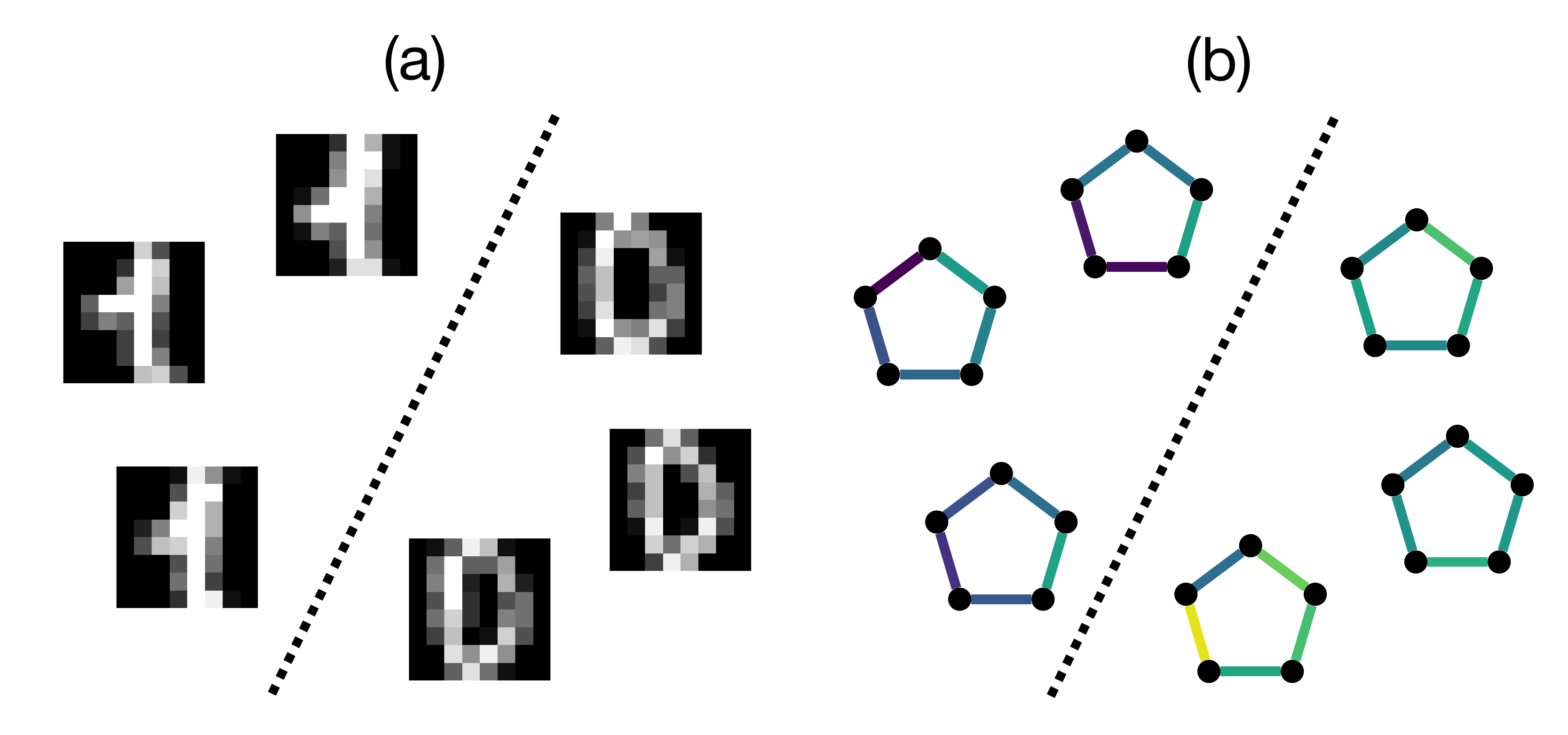}
    \caption{\BLUE{Diagrams illustrating the two classification tasks that are the focus of the current research: (a) classification of small-scale binary images, (b) classification of weighted graphs (here weights are mapped to colors) based on the minimal energy of their Ising Hamiltonians.}}
    \label{fig:intro}
\end{figure}

\BLUE{Here we present the results of benchmarking a quantum processor based on trapped $^{171}$Yb$^{+}$ ions with the use of up to five qubits.
The study involves implementing a quantum-enhanced support vector machine (SVM) algorithm~\cite{Havlek2018SupervisedLW,Schuld2021qmodels} for the two classification tasks shown in Fig.~\ref{fig:intro}.
The first task is a supervised binary classification of small-scale digit images from a publicly available dataset.
The images are intentionally chosen so that they can be classified with $100\%$ accuracy.
For this task, we consider different types of quantum encodings of the dataset and several degrees of transpilation optimizations for corresponding quantum circuits.
For each quantum encoding, we obtain a classifier that is 100\% accurate on both training and test sets.
The second task is a supervised binary classification of weighted graphs with ring topology, according to the spectrum of their corresponding Ising Hamiltonians.
For this purpose we synthetically generate a dataset with $n$-vertex graphs with $n=3,4,5$.
The idea of this task is to benchmark the ability of quantum processors to embed a $2^n$-element spectrum of the Ising Hamiltonian into the phases of complex coefficients of $n$-qubit superposition state through the application of $n$ entangling operations.
We would like to point out that our approach diverges from previously proposed schemes for using quantum boson samplers to address the graph classification problem~\cite{schuld2020measuring}.
We investigate the impact of transpilation optimization on the corresponding circuits and find that, despite the detrimental effect of non-optimized transpiling on the quality of reconstructed kernel matrices, we are able to obtain a classifier with 100\% accuracy on both the training and testing sets for both non-optimized and optimized transpilation scenarios.}
Our results demonstrate that the quantum processor can correctly solve the basic, small-scale classification task\BLUE{s} considered; as we expect, with the increase of their capabilities quantum processors have the potential to become a useful tool for machine learning applications.

\section{Quantum-enhanced SVM}

In this section, we briefly revisit the fundamental concepts underlying the binary classification using the quantum-enhanced SVM.
For more detailed information, we refer the reader to Refs.~\cite{Havlek2018SupervisedLW,Schuld2021qmodels}.
Let $\{({\bf x}_i, l_i)\}_{i=1}^L$ denote an $L$-element labeled training dataset, consisting of feature vectors ${\bf x}_i \in \mathbb{R}^d$, where $d$ is the dimensionality of the feature space, and corresponding labels $l_i=\pm 1$.
Within the standard SVM framework, predicting a label $l'$ for a test sample ${\bf x}'$ can be formulated as:
\begin{equation} \label{eq:pred}
    \begin{aligned}
        l'&:={\rm sign}\left( K({\bf x'}, {\bf w})+b\right)\\
        &={\rm sign}\left(\sum_{s\in S}\alpha_s l_s K({\bf x}', {\bf x}_s) +b\right)\\
    \end{aligned}
\end{equation}
with ${\bf w}=\sum_{s\in S}\alpha_s l_s {\bf x}_s$.
Here $K(\cdot,\cdot):\mathbb{R}^d\times \mathbb{R}^d \rightarrow \mathbb{R}$ represents a \emph{kernel function}, which is a prespecified function that provides positive semidefiniteness property to every matrix $\left[K({\bf y}_i, {\bf y}_j)\right]_{ij}$ constructed based on an arbitrary finite set of feature vectors $\{{\bf y}_i\}$;
the subset $S \subset \{1,\ldots,L\}$ is a collection of indices of \emph{support vectors} taken from the training dataset; and $\alpha_s$, $b$ are real numbers.
The values of $S$, $\{\alpha_s\}$, and $b$ come from the training process.
More specifically, $\{\alpha_i\}_{i=1}^L$ appears as the solution to the quadratic programming optimization problem
\begin{equation} \label{eq:optproblem}
    \begin{aligned}
        \frac{1}{2} \sum_{i,j=1}^L \alpha_i\alpha_jl_il_j K({\bf x}_i, {\bf x}_j) - \sum_{i=1}^L\alpha_i \rightarrow \min\quad{\rm s.t.}\\
        0 \leq \alpha_i \leq C {\rm~for~} i=1,\ldots,L, \quad \sum_{i=1}^L\alpha_il_i=0,
    \end{aligned}
\end{equation}
where $C>0$ is a regularization parameter; the set of support vector indices $S$ is defined as a subset of $i\in\{1,\ldots,L\}$ with $\alpha_i>0$; and 
\begin{equation} \label{eq:bconst}
    b:= \frac{1}{|S|}\sum_{s\in S}\left(l_s-\sum_{m\in S}\alpha_ml_mK({\bf x}_m,{\bf x}_s)\right),
\end{equation}
where $|S|$ is the number of elements in $S$.
Intuitively, as a result of the training process, the entire space of feature vectors ${\bf x} \in \mathbb{R}^d$ is divided into two regions: 
\begin{equation} \label{eq:classification_rule}
	K({\bf x}, {\bf w})+b \geq 0 \quad \text{ and } \quad K({\bf x}, {\bf w})+b <0, 
\end{equation}	
corresponding to labels of $+1$ and $-1$, respectively. 
The border of the division is specified by $\mathbf{w}$ and $b$ that are determined by a limited set of support vectors with their labels $\{({\bf x}_s, l_s)\}_{s \in S}$ and coefficients $\{\alpha_s\}_{s\in S}$. 
The crucial fact is that the optimization problem~\eqref{eq:optproblem} can be efficiently solved using standard quadratic optimization methods in polynomial time, provided that the elements of the \emph{kernel matrix} 
$K_{ij}:=K({\bf x}_i, {\bf x}_j)$ are precomputed.

The elements of the kernel matrix $K({\bf x},{\bf y})$ can be interpreted as an inner product $\phi({\bf x})\cdot \phi({\bf y})$ of some vectors $\phi({\bf x})$ and $\phi({\bf y})$.
These vectors may have a dimension that is significantly different from the original feature space.
During the training process, defined by Eq.~\eqref{eq:optproblem} and Eq.~\eqref{eq:bconst}, 
and prediction  described by Eq.~\eqref{eq:pred},  
only scalar products of the form $K({\bf x}_i,{\bf x}_j)$ and $K({\bf x}_s,{\bf x}')$ are required [there is no need to explicitly compute the $\phi(\cdot)$ vectors].
The concept of quantum-enhanced SVM is based on the design of a kernel function, $K(\cdot,\cdot)$, that can be calculated on a quantum computer. 
This kernel function corresponds to the embedding of the original feature vectors into a Hilbert space of quantum states. 
In this scenario, 
the mapping from feature vectors to their corresponding quantum states may take on a quite complex form due to the creation of entangled quantum states during the calculation of individual kernel elements. 
In the next section a specific implementation of the considered quantum-enhanced SVM algorithm is provided.

\section{Supervised classification of digits images}

Here we consider a task of digits images supervised binary classification with the use of the quantum-enhanced SVM approach.
In particular, we consider a small subset of images of handwritten digits 0 and 1 from the Optdigits dataset~\cite{optdigits}.
Training and test datasets of digits images are shown in Fig.~\ref{fig:datasets}.
The training set contains six images, the test set contains four images. 
All images are naturally divided into two classes, depicting zeros and ones, respectively. 
The size of each image is $8\times8$ pixels. 

The process of the kernel matrix estimation is performed with the following quantum circuit:
\begin{equation} \label{eq:whole_circ}
\begin{adjustbox}{width=0.5\linewidth}
\begin{quantikz}\Large
\lstick{\ket{0}} &\gate[4]{U(\mathbf{x})}&\gate[4]{U^{\dagger}(\mathbf{y})}& \meter{} \\
\lstick{\ket{0}} &&& \meter{} \\
\lstick{\ket{0}} &&& \meter{} \\
\lstick{\ket{0}} &&& \meter{} \\
\end{quantikz}
\end{adjustbox}
\end{equation}

Here $U(\mathbf{x})$ stands for an embedding circuit, transforming an initial state $\ket{0}^{\otimes N}$ into a state $\ket{\phi(\mathbf{x})}$, corresponding to the image $\mathbf{x}$. Here $N$ denotes the number of used qubits. It is followed by the action of $U^\dagger(\mathbf{y})$ and terminated by the measurement. The kernel matrix element 
\begin{equation}
    \begin{aligned}
        K(\mathbf{x}, \mathbf{y}) &= |\langle \phi(\mathbf{x}) | \phi(\mathbf{y}) \rangle |^2
        \\&= \left(\bra{\phi({\bf x})}\otimes \overline{\bra{\phi({\bf x})}}\right)
        \left(
            \ket{\phi({\bf y})}\otimes\overline{\ket{\phi({\bf y})}}
        \right),
    \end{aligned}
\end{equation}
which is the fidelity between states $\ket{\phi(\mathbf{x})}$ and $\ket{\phi(\mathbf{y})}$,
is then given by the probability of finding the quantum register back in the state $\ket{0}^{\otimes N}$ (here, the overline denotes a complex conjugation).

To allow SVM algorithm to build $100\%$ accurate classifier, the encoding circuit should be chosen in such a way, that embedded points of different classes (images of zeros and ones) are linearly separable. That is, there must be some hyperplane in the 
\GREEN{doubled} 
Hilbert space which separates all points from the one class from another. 
In this \BLUE{section} we consider three real-valued quantum encoding methods, which differ in terms of the number of qubits used and the number of gates employed (especially two-qubit entangling gates). This allows us to investigate the stability of the quantum-enhanced SVM algorithm \BLUE{implementation} on an existing quantum processor.

As the first preprocessing step for all three encodings, we scale pixel intensities of all considered digits images to values between 0 and 1. For simplicity, we use only the central pixels of the digits images as they provide enough information for classification. For the first two encodings, we select from each image the intensities of pixels with coordinates $(3,3)$, $(3,4)$, $(4,3)$, and $(4,4)$, multiply them by $\pi$ and flatten them as a vector $\mathbf{x}=(x_1,x_2,x_3,x_4)$. 

The first and the simplest encoding consists of a layer of $R_Y$ gates on four qubits, where $R_Y$ stands for a standard rotation around $Y$ axis on the Bloch sphere:
\begin{equation}\label{eq:enc1}
\begin{adjustbox}{width=0.5\linewidth}
\begin{quantikz}\Large
&\gate[4]{U_1(\mathbf{x})} &\midstick[4, brackets=none]{=} &\gate{R_Y(x_1)}& \\
&&&\gate{R_Y(x_2)}& \\
&&&\gate{R_Y(x_3)}& \\
&&&\gate{R_Y(x_4)}& \\
\end{quantikz}
\end{adjustbox}
\end{equation}

The kernel matrix elements for this encoding in the absence of noise are given by the relation
\begin{equation} \label{eq:ker12}
	K(\mathbf{x}, \mathbf{y}) = \prod\limits_{i=1}^4\cos^2{\left(\frac{x_i-y_i}{2}\right)}.
\end{equation}
Due to the lack of entangling operations, we can anticipate that the results of using this encoding on a real processor will closely resemble the ideal, noise-free case.

In the second encoding, to study the effect of the two-qubit gates noise on the classification accuracy, we add another layer of two \BLUE{${\sf CX}$ (controlled-NOT)} gates: 
\begin{equation}\label{eq:enc2}
\begin{adjustbox}{width=0.68\linewidth}
\begin{quantikz}[column sep = 2em]\Large
&\gate[4]{U_2(\mathbf{x})} &\midstick[4, brackets=none]{=} &\gate{R_Y(x_1)}& \ctrl{1} & \\
&&&\gate{R_Y(x_2)}& \targ{}  & \\
&&&\gate{R_Y(x_3)}& \ctrl{1} & \\
&&&\gate{R_Y(x_4)}& \targ{}  & \\
\end{quantikz}
\end{adjustbox}
\end{equation}
We note that due to the mirror structure of the circuit~\eqref{eq:whole_circ} used for computing kernel elements, in the noiseless case, the resulting values for kernel elements have to be the same as for the first encoding method and given by Eq.~\eqref{eq:ker12}.
However, making ${\sf CX}$ gates brings additional noises to observed values, and this is the way how it provides a possibility for benchmarking the noise stability of the algorithm.

The third embedding relies on the amplitude encoding approach~\cite{Schuld2018supervised, Mottonen2005Transformation}. Its idea is to prepare a quantum state with amplitudes that are equal to the components of a given unit Euclidean vector $\mathbf{\overline{x}}$. An advantage of such embedding is that due to the usage of entanglement it allows one to encode more information in the same number of qubits. Namely, a unit vector with $2^N$ components can be encoded using $N$ qubits. Here we use $N=2$ qubits to encode values of three central pixels with coordinates $(3,3)$, $(3,4)$, and $(4,4)$. Their intensities scaled to values between 0 and 1 and padded with 0.25 form a vector $\mathbf{x}=(x_1,x_2,x_3,0.25)$. As amplitude encoding can be applied only to unit vectors, we normalize it: $\mathbf{\overline{x}}=\mathbf{x}/|\mathbf{x}|$. The padding prevents the case where all components of $\mathbf{x}$ are zero and normalization is impossible. Afterwards we calculate rotation angles $\mathbf{a} = (a_1, a_2, a_3, a_4, a_5)$ for the gates in the following amplitude encoding circuit:
\begin{equation}\label{eq:enc3}
\resizebox{0.88\columnwidth}{!}{
\begin{quantikz}[column sep = 0.5  em]\Large
&\gate[2]{U_3(\mathbf{x})} &\midstick[2, brackets=none]{=} &\gate{R_Y(a_1)}& \ctrl{1} &  & \ctrl{1} & \gate{R_X(\pi)} & \ctrl{1} &  & \ctrl{1} & \gate{R_X(\pi)} & \\
&&&               & \targ{}  & \gate{R_Y(a_2)} & \targ{} & \gate{R_Y(a_3)} & \targ{} & \gate{R_Y(a_4)} & \targ{} & \gate{R_Y(a_5)} & \\
\end{quantikz}
}
\end{equation}
The rotation angles $\mathbf{a}$ are calculated as follows~\cite{Schuld-vctut}:
\begin{equation}\label{eq:amp_formulas}
\begin{array}{l} 
\mathbf{a} := (\beta_2, -\beta_1 / 2, \beta_1 / 2, -\beta_0 /2 , \beta_0 / 2 ), \\[3pt]
\beta_0 := 2 \arcsin{(\overline{x}_2 / \sqrt{\overline{x}_1^2 + \overline{x}_2^2 + \varepsilon})}, \\[3pt]
\beta_1 := 2 \arcsin{(\overline{x}_4 / \sqrt{\overline{x}_3^2 + \overline{x}_4^2 + \varepsilon})}, \\[3pt]
\beta_2 := 2 \arcsin{(\overline{x}_3^2 + \overline{x}_4^2)}.
\end{array}
\end{equation}
Here, vector $\mathbf{\overline{x}}=(\overline{x}_1, \overline{x}_2, \overline{x}_3, \overline{x}_4)$ is a result of the $\mathbf{x}$ normalization and $\varepsilon = 10^{-12}$ is used to prevent division by zero. We note that circuit \eqref{eq:enc3} and formulas \eqref{eq:amp_formulas} are valid only when all components of $\mathbf{\overline{x}}$ are non-negative, which is the case here.
The kernel matrix in the absence of noise here is given by
\begin{equation} \label{eq:ker3}
    K(\mathbf{x}, \mathbf{y}) = |\langle \mathbf{\overline{x}} | \mathbf{\overline{y}}) \rangle |^2.
\end{equation}
From the circuit, given by expression~\eqref{eq:enc3}, it can be seen that the cost of a more dense information encoding is the increased number of two-qubit operations, which usually are the most noisy elements in all quantum algorithms.  Therefore, the third encoding can be used for studying the noise robustness of embedding classical information in entangled quantum states within quantum-enhanced SVM algorithm running on a real quantum processor.

In what follows, we refer to these encodings as (i) $R_Y$, (ii) $R_Y+{\sf CX}$, and (iii) amplitude encodings, correspondingly. 
Analytically, all three encodings send vectors $\mathbf{x}$, which correspond to various classes, to different areas in the quantum state space, 
thus making the images of each embedding linearly separable and allowing one to use the SVM algorithm for constructing $100\%$ accurate separating hyperplane in each case.

After estimating the kernel matrices using the quantum computer, we employ the classical implementation of the SVM algorithm from the Scikit-learn toolkit~\cite{scikit-learn}. 
We use default values of the parameters of the algorithm, in particular, the regularization parameter $C=1$. 
\BLUE{The accuracy of $100\%$} on both training and test sets in the experiments using a quantum emulator for each quantum encoding is obtained. 

\BLUE{
\section{Supervised classification of weighted graphs}

Here we consider a supervised classification of weighted graphs according to the spectrum of their corresponding Ising Hamiltonians.
Consider a $n$-vertex undirected weighted graph specified by an $n\times n$ weight matrix ${\bf g}=(g_{jk})$ with $g_{jj}=0$ and $g_{jk}=g_{kj}$ ($g_{ij}$ is a weight of an edge between $i$-th and $j$-th node).
Each such graph ${\bf g}$ can be associated with an $n$-qubit Ising Hamiltonian
\begin{equation} \label{eq:graph_Hamiltonian}
    {\cal H}_{\bf g}=\frac{1}{2}\sum_{i,j}g_{jk}\sigma_z^j\otimes\sigma_z^k,
\end{equation}
which is a diagonal $2^n$ matrix.
Hereinafter, $\sigma_{\alpha}^l$ with $\alpha\in\{x,y,z\}$ stands for standard Pauli-$\alpha$ operator acting on the $l$-th qubit.
The eigenbasis of ${\cal H}_{\bf g}$ can be taken as a set of computational basis states, where each state $\ket{x}$ with $x\in\{0,1\}^n$ has an energy (eigenvalue) $E_x$.
Note that in accordance with the structure of~\eqref{eq:graph_Hamiltonian}, $E_x=E_{{\rm inv}(x)}$, where ${\rm inv}(x)$ is an all-bits flipped version of $x$.
In what follows we consider a graph classification problem with respect to states in the bottom of the spectrum.
We recall that finding ground state of an Ising Hamiltonian can be reduced to the MAX-CUT problem, known to be NP-hard.

The classification problems are formulated as follows. 
For $n = 3, 4, 5$, we generate training and test sets consisting of $L_{\rm train} = 20$ and $L_{\rm test} = 10$ ring-topology graphs, respectively.
To generate each graph, we first sample $n$ weights $g_{i, (i + 1) \text{mod} n}$ (here $i=0,\dots,n-1$) from a normal distribution with zero mean and unit standard deviation. 
We then check whether the resulting graph can be labeled with one of the two labels ($\pm 1$), according to the classification rules specified in Table~\ref{tab:classification_rules}. 
If so, we include the graph in our training or test set, otherwise we repeat the weight sampling process.

\begin{table}[]
    \centering
    \begin{tabular}{c|c|c}
        $n$ & Label $+1$ & Label $-1$\\ \hline
        $3$ & $\{\ket{000},\ket{111}\}$ & \{\ket{100},\ket{011},\ket{010},\ket{101},\ket{001},\ket{001}\}  \\
        $4$ & $\{\ket{0000},\ket{1111}\}$ & $\{\ket{0101},\ket{1010}\}$ \\
        $5$ & $\{\ket{00000},\ket{11111}\}$ & $\{\ket{01010},\ket{10101}\}$
    \end{tabular}
    \caption{The principle of generating datasets for graph classification problems:
    A given $n$-vertex graph ${\bf g}$ is assigned with a particular label if the set of $s$ levels with the lowest energies of the Hamiltonian ${\cal H}_{\bf g}$ coincides with the corresponding $s$-element set from the table. We note that due to randomized graph, generation, a probability to obtain degeneracy of levels (other from $E_x=E_{{\rm inv}(x)}$ is negligible.}
    \label{tab:classification_rules}
\end{table}

The quantum-enhanced SVM for solving these problems is based on embedding a graph ${\bf g}$ in a quantum state $\ket{\psi({\bf g})} = V({\bf g})\ket{0}^{\otimes n}$ with the encoding operator
\begin{equation} \label{eq:graph-encodings}
    V({\bf g})=\prod_{j,k>j}\exp\left(-i\gamma \widetilde{g}_{jk} \sigma_x^j\otimes\sigma_x^k\right),
\end{equation}
where $\widetilde{g}_{jk}:=g_{jk}/\max_{j'k'}|g_{j'k'}|$.
Note that in contrast to~\eqref{eq:graph_Hamiltonian}, $V({\bf g})$ is diagonal in the Hadamard basis.
In the case of $n=3$ qubit, this operator can be depicted as
\begin{equation}
\resizebox{0.88\columnwidth}{!}{
\begin{quantikz}
&\gate[wires=3]{V({\bf g})}& \midstick[3,brackets=none]{=} &\gate[wires=2]{{\sf MS}(\gamma \widetilde{g}_{01})}&&\gate{{\sf MS}(\gamma \widetilde{g}_{02})}\wire[d][2]{q}&\\
&&&&\gate[wires=2]{{\sf MS}(\gamma \widetilde{g}_{12})}&&\\
&&&&&\gate{{\sf MS}(\gamma \widetilde{g}_{02})}&
\end{quantikz},    
}
\end{equation}
where ${\sf MS}(\chi)=\exp\left(-i\chi\sigma_x\otimes\sigma_x\right)$ denotes a M\o{}lmer-S\o{}rensen (MS) gate, discussed in the next section.
We also note that the embedding~Eq.~\eqref{eq:graph-encodings} is not limited to cyclic graphs alone.
We set the parameter $\gamma$ to 0.8, as we found that this produces a stable performance of the resulting classifiers for all problems under consideration.
To understand the intuition behind the chosen encoding, we can rewrite $\ket{\psi(\mathbf{g})}$ as:
\begin{equation}
    \begin{aligned}
        \ket{\psi({\bf g})} &= V({\bf g})\ket{0}^{\otimes n} \\&= H^{\otimes n} \exp\left({-i\frac{\gamma}{g_{\max}}}{\cal H}_{\bf g}\right) H^{\otimes n} \ket{0}^{\otimes n}\\
        &=H^{\otimes n}\sum_{x\in\{0,1\}^n}\frac{1}{\sqrt{2^n}}\exp\left(-i \frac{\gamma}{g_{\max}} E_x\right)\ket{x}
    \end{aligned}
\end{equation}
where $H$ is the Hadamdard gate and $g_{\max}:=\max_{jk}|g_{jk}x|$.
One can see that $V({\bf g})$ maps $2^n$ eigenvalues $E_x$ of ${\cal H}_{\bf g}$ into the phases of the symmetrical superposition state, using $n$ MS gates.
The design of this coding scheme is based on the architecture of the single-layer quantum approximate optimization algorithm (QAOA)~\cite{farhi2014quantum}. 
However, unlike the original QAOA, we do not optimize parameters in any kind of a feed-back loop. 
Additionally, we omitted the mixing layer due to the symmetrical nature of the employed circuit for kernel estimation, as illustrated in~\eqref{eq:whole_circ}.

The elements of the kernel matrix for each pair of graphs, ${\bf g}$ and ${\bf g}'$, are calculated as the overlap between the respective state vectors, as was done in the previous case:
\begin{equation} \label{eq:kernel-for-graphs}
    \begin{aligned}
        K({\bf g}',{\bf g})&=|\langle \psi({\bf g}')\ket{\psi({\bf g})}|^2\\
        &=|\bra{0}^{\otimes n}V({\bf g}')^\dagger V({\bf g})\ket{0}^{\otimes n}|^2\\
        &=|\bra{0}^{\otimes n}\widetilde{V}({\bf g}',{\bf g})\ket{0}^{\otimes n}|^2,\\
        \widetilde{V}({\bf g},{\bf g}')&=  \prod_{j,k>j}\exp\left(-i\gamma (\widetilde{g}_{jk}-\widetilde{g}_{jk}') \sigma_x^j\otimes\sigma_x^k\right).
    \end{aligned}
\end{equation}
This overlap can be computed as probability of getting all-zeros outcome after running the corresponding circuit.
We note that merging the product of $V({\bf g})$ and $V({\bf g}')^\dagger$ into $\widetilde{V}({\bf g}',{\bf g})$ reduces the number of entangling MS operations required.

Running the algorithm for the generated datasets with regularization parameter $C=1$ and using an ideal state-vector simulator results in 100\% classification accuracy both for training and test sets for all $n=3,4,5$.
In what follows, we describe the experiments that have been performed using the trapped-ion quantum processor.
}

\section{Trapped-ion quantum processor}

Ytterbium ($^{171}$Yb$^{+}$) ions are actively used in quantum technologies, specifically, metrology~\cite{huntemann2012high, khabarova2022toward, Schwindt2016} and computing~\cite{Sage2019, moses2023race, chen2023benchmarking, Zalivako2024}.
This species possesses a convenient energy structure, which allows one to use its efficient and ensure relatively simple laser cooling~\cite{Zalivako2019b}, state initialization and readout~\cite{Ejtemaee2010, semenin2021optimization}. 
Rich level structure also provides several ways of qubits encoding, such as in hyperfine sublevels of the ground state $^2S_{1/2}$ (microwave qubit)~\cite{wang2021single, Foss-Feig2021, Monroe2019} 
or in the states coupled by a narrow E2 optical transition $^2S_{1/2}\to\,^2D_{3/2}$ (optical qubit)~\cite{Zalivako2024}. 

Our quantum processor (its detailed description can be found in Refs.~\cite{Zalivako2024, zalivako2025quantum}) is based on a string of 10 $^{171}$Yb$^{+}$ ions inside a linear Paul trap. 
The trap secular frequencies are $\{\omega_x, \omega_y, \omega_z\}=2\pi\times\{3.650, 3.728, 0.129\}$\,MHz. The qubits are encoded in states $\ket{0}=\,^2S_{1/2}(F=0,m_F=0)$ and $\ket{1}=\,^2D_{3/2}(F=2,m_F=0)$ 
coupled by an optical transition at 435.5\,nm with an upper state lifetime of $\tau=53$\,ms. 
Our processor also supports the qudit regime~\cite{Kolachevsky2022,Zalivako2024}, where other Zeeman sublevels of the upper state are used for information encoding as well.
Although qudit encoding may provide various advantages, in particular, for decompositions of multiqubit gates~\cite{Ralph2007,White2009,Ionicioiu2009,Wallraff2012,Gokhale2019,Nikolaeva2022}, 
below we focus on the qubit regime with states $\ket{0}$ and $\ket{1}$.

Before each experimental shot, the ion chain is cooled to the ground state along trap axes $x$ and $y$, and all ions are initialized to the $\ket{0}$ state by optical pumping. 
After that, quantum gates are applied to the qubits. 
Quantum gates are performed by applying laser pulses at 435.5\,nm to the ions. 
For this purpose, the setup is equipped with two addressing beams, which can be scanned along the ion chain using acousto-optical deflectors and individually interact with particular qubits. 

As single-qubit gates the system supports 
\begin{equation}
R_{\phi}(\theta) = \exp(-i\sigma_{\phi}\theta/2),  
\end{equation}
where $\sigma_\phi = \sigma_x\cos\phi  + \sigma_y\sin\phi $, $\sigma_x$ and $\sigma_y$ are standard Pauli matrices, and $\phi, \theta$ are arbitrary real angles. The fidelity of such operation is \BLUE{$99.95\%$}, measured using randomized benchmarking. As a two-qubit gate, we use the ${\sf MS}(\chi)$ gate~\cite{Blatt2003-2,Molmer-Sorensen1999,Molmer-Sorensen1999-2,Molmer-Sorensen2000} with a variable rotation angle $\chi$.
\BLUE{The mean fidelity of fully-entangling ${\sf MS}(\pi/4)$ gate across all possible pairs in 10-qubit quantum register} was estimated to be \BLUE{96.3\%} by measuring parity oscillations after Bell state preparation \BLUE{and after correcting results for state preparation and measurement errors (SPAM). The latter amount to approximately $1\%$ per ion}. The \BLUE{gate} fidelity at the moment is limited by the long duration of the two-qubit gate (approximately 1~ms) relatively to the upper qubit state lifetime of 53 ms. \BLUE{We also note, that the ${\sf MS}(\chi)$ gate error monotonically increases with $\chi$, however is not fully proportional to it.} The dephasing time of the qubits is 30 ms and is limited by the laser frequency stability. 

The gates sequence is followed by the readout procedure implemented using electron shelving technique~\cite{Leibfried2003, semenin2021optimization}. 

As it can be seen from expressions~\eqref{eq:whole_circ}, \eqref{eq:enc1}, \eqref{eq:enc2}, and \eqref{eq:enc3}, our circuits \BLUE{for digit images classification} contain only single-qubit rotations, which are included in the native gate set, 
and ${\sf CX}$ gates. \BLUE{The latter we transpile via the native gates supported by the processor as}:
\begin{equation}\label{eq:cx-ms}
\begin{adjustbox}{width=0.88\linewidth}
\begin{quantikz}
&\ctrl{1}&\midstick[2, brackets=none]{=} &\gate{R_Y(-\frac{\pi}{2}}&\gate[wires=2]{{\sf MS}(\frac{\pi}{4})}&\gate{R_X(-\frac{\pi}{2})}&\gate{R_Y(\frac{\pi}{2})}&\\
&\targ{} & & &&\gate{R_X(\frac{\pi}{2})}&&
\end{quantikz}
\end{adjustbox}
\end{equation}
Here $R_X(\theta):=R_0(\theta)$ and $R_Y(\theta):=R_{\pi/2}(\theta)$. \BLUE{In experiments aimed for weighted graphs classification all the used gates ${\sf MS}(\chi)$ already included in the native gate set.}

\section{Results}
We summarize the experimental results for classifying digits and graphs in Tables~\ref{tab:results} and \ref{tab:results-graphs}, correspondingly. 
For a more detailed characterization of the results, in these tables \BLUE{along with classification accuracies} we also provided distances between kernel matrices, obtained experimentally and calculated with an ideal emulator.
Here we define distance as  
\begin{equation} \label{eq:dist-func}
    d(A,B) = \max_{ij}|A_{ij}-B_{ij}|,    
\end{equation}
where $A$ and $B$ are matrices to compare.

\begin{table}[]
\setlength\extrarowheight{2pt}
\begin{tabular}{|cc|cc|cc|}
\hline
\multicolumn{2}{|c|}{}                    & \multicolumn{2}{c|}{
\begin{minipage}[t]{0.15\textwidth}
Non-optimized transpilation
\end{minipage}
} & \multicolumn{2}{c|}{
\begin{minipage}[t]{0.15\textwidth}
Optimized transpilation
\end{minipage}
} \\ \cline{1-6} 
\multicolumn{1}{|c|}{Enc.}                    & Shots                                     & \multicolumn{1}{c|}{Training}  & Test& \multicolumn{1}{c|}{Training}  & Test  \\ \hline
\multicolumn{1}{|c|}{\multirow{3}{*}{$R_Y$}} & 2048 & \multicolumn{1}{c|}{100 (0.09) }        & 100 (0.08)     & \multicolumn{1}{c|}{-}        & -       \\ \cline{2-6} 
\multicolumn{1}{|c|}{}                            & 2048 & \multicolumn{1}{c|}{100 (0.06)}        & 100 (0.06)      & \multicolumn{1}{c|}{-}        & -       \\ \cline{2-6} 
\multicolumn{1}{|c|}{}                            & 2048 & \multicolumn{1}{c|}{100 (0.07)}        & 100 (0.05)      & \multicolumn{1}{c|}{-}        & -      \\ \hline
\multicolumn{1}{|c|}{\multirow{3}{*}{$R_Y+{\sf CX}$}}      &1024                     & \multicolumn{1}{c|}{100 (0.32) }        & 100 (0.32)      & \multicolumn{1}{c|}{100 (0.08)}        & 100 (0.09)       \\ \cline{2-6} 
\multicolumn{1}{|c|}{}                            & 1024 & \multicolumn{1}{c|}{100 (0.52)}         & 100 (0.35)      & \multicolumn{1}{c|}{100 (0.04)}        & 100 (0.05)  \\ \cline{2-6}   
\multicolumn{1}{|c|}{}                            & 1024 & \multicolumn{1}{c|}{100 (0.24) }         & 100 (0.24)      & \multicolumn{1}{c|}{100 (0.06)}        & 100 (0.02) \\ \hline
\multicolumn{1}{|c|}{\multirow{2}{*}{Ampl.}} &  1024 & \multicolumn{1}{c|}{100 (0.45)}        & 100 (0.48)      & \multicolumn{1}{c|}{100 (0.44)}        & 100 (0.6)       \\ \cline{2-6} 
\multicolumn{1}{|c|}{}                            & 1024 & \multicolumn{1}{c|}{83 (0.53)}         & 100 (0.6)     & \multicolumn{1}{c|}{100 (0.5)}        & 100 (0.43)      \\ \hline
\end{tabular}
\caption{Accuracy of digits' classifiers (in percentages) that were obtained in all experiments on training and test datasets with all encodings for both optimized and non-optimized transpilations. 
In brackets, distances~\eqref{eq:dist-func} between ideal and experimentally reconstructed kernel matrices are shown.}
\label{tab:results}
\end{table}

\begin{table}[]
\begin{tabular}{|p{6mm}|cc|cc|}
\hline
\multirow{2}{*}{$n$} & \multicolumn{2}{c|}{Non-optimized transpilation} & \multicolumn{2}{c|}{Optimized transpilation} \\ \cline{2-5} 
                     & \multicolumn{1}{c|}{Training}     & Test         & \multicolumn{1}{c|}{Training}   & Test       \\ \hline
3                    & \multicolumn{1}{c|}{100 (0.36)}   & 100 (0.31)   & \multicolumn{1}{c|}{100 (0.13)} & 100 (0.10) \\ \hline
4                    & \multicolumn{1}{c|}{100 (0.28)}   & 100 (0.51)   & \multicolumn{1}{c|}{100 (0.08)} & 100 (0.12) \\ \hline
5                    & \multicolumn{1}{c|}{100 (0.63)}   & 100 (0.59)   & \multicolumn{1}{c|}{100 (0.06)} & 100 (0.08) \\ \hline
\end{tabular}
\caption{\BLUE{Accuracy of graphs' classifiers (in percentages) for various values of vertex number $n$ within non-optimized and optimized transpilations.
The number of shots for estimating each kernel element is equal to 1024 for all experiments.
As in Table~\ref{tab:results}, distances~\eqref{eq:dist-func} between ideal and experimentally reconstructed kernel matrices are shown in brackets.}}
\label{tab:results-graphs}
\end{table}

\begin{figure}
    \centering
    \includegraphics[width=0.8\linewidth]{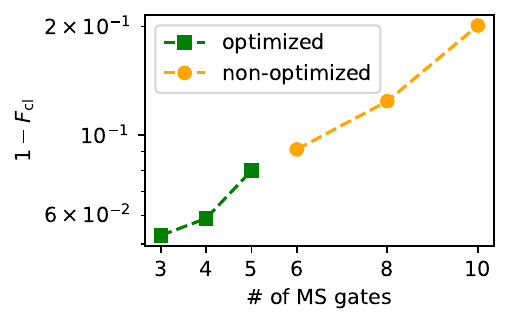}
    \caption{Average classical infidelities of circuits in graphs classification experiments as number of parametrized MS gates employed ($n$ for optimized and $2n$ for non-optimized transpilation modes).}
    \label{fig:fidelity}
\end{figure}

\BLUE{Within the digits' classification, we have studied algorithm performance and robustness with two transpilation schemes, which we further refer to as ``non-optimized transpilation'' and ``optimized transpilation''. In the ``non-optimized transpilation'' case the circuits were implemented exactly as they are shown in Eqs. ~\eqref{eq:whole_circ}, \eqref{eq:enc1}, \eqref{eq:enc2}, and \eqref{eq:enc3}. 
For ``optimized transpilation'' we take the transpiled circuits we described above and apply the following optimizations: (i) combine several single-qubit rotations around the same axis into one rotation; 
(ii) eliminate single-qubit rotations where, after combining, rotation angle is a multiply of $2\pi$; and (iii) eliminate two consecutive $\sf{CX}$ gates. 
Such actions significantly reduce number of both single- and two-qubit operations and, therefore, noise level.}

For the first $R_Y$ encoding, we performed three separate experiments on the same dataset and trained three classifiers. As the difference between optimized and non-optimized transpilations here is negligible, we performed these experiments only for non-optimized circuits. 
In each experiment, 2048 shots were made for every circuit.
The accuracy of each classifier was obtained to be equal to $100\%$ on both training and test sets. 

For the second $R_Y(\theta)+{\sf CX}$ encoding, we ran the experiment three times for both optimized and non-optimized transpilations. 
In total, we performed six experiments on the same dataset and trained six classifiers with 1024 shots used for every circuit in each experiment.
As for the first $R_Y$ encoding, the accuracy of each classifier was $100\%$ in all cases.

For amplitude encoding, we ran the experiment two times for both optimized and non-optimized transpilations. In this case, 1024 shots for each circuit were performed.
In total, four experiments on the same dataset and training of four classifiers were performed.
Among executed quantum circuits, this encoding is the most hardware-demanding as each circuit consists in the non-optimized case of eight two-qubit gates.
Therefore, due to the noise, the achieved accuracy of one classifier on the training set appeared to be 83\%.
However, in all other experiments with amplitude encoding, the achieved accuracy of each classifier is 100\%. 

\BLUE{
To address the problem of graphs classification, we conducted a total of six experiments: for each dataset ($n=3, 4, 5$), we considered both non-optimized and optimized transpilation modes.
In the former case, we sequentially applied gates $V({\bf g})$ and $V({\bf g'})^\dagger$ to the qubits initialized in $\ket{0}^{\otimes n}$ which resulted in $2n$ entangling MS gates in the circuit. 
In the latter, we used $n$ MS gates in total to realize a joint $\widetilde{V}({\bf g}', {\bf g})$ gate [see Eq.~\eqref{eq:kernel-for-graphs}]. For each quantum circuit 1024 shots were made.

We observed (see Table~\ref{tab:results-graphs}), that despite the obvious loss in accuracy of non-optimized tanspilation compared to the optimized one, the classifier successfully copes with a problem for all considered values of $n$ with $100\%$ accuracy both on the training and test datasets.

To further verify the experimental results, we computed the classical fidelity
\begin{equation}
    F_{\rm cl}=\sum_{x\in\{0,1\}^n}\sqrt{P_{\rm exp}(x)P_{\rm ideal}(x)},
\end{equation}
for all circuits run within the graph classification task. 
Here $P_{\rm exp}(x)$ and $P_{\rm ideal}(x)$ are the experimentally observed and expected probability distributions of obtaining bitstring $x$, respectively. 
The results of averaging these fidelities over all circuits with the same number of MS gates used are shown in Fig.~\ref{fig:fidelity}.
One can observe a clear exponential increase in the infidelity $1-F_{\rm cl}$, which confirms the expectation that the performance of the circuit implementation degrades with the number of entangling gates used.
At the same time, we can see a slight improvement in the accuracy of kernel matrix reconstruction in optimally transpiled circuits with increasing $n=3,4,5$ in Table~\ref{tab:results-graphs}, which can be explained by a non-trivial interplay between noise effects caused by MS gates and a bias in readout measurements.
}

\section{Discussion}
The main target parameter in the considered tasks is the classification accuracy on both training and test datasets.
The $100\%$ classification accuracy achieved in all experiments can be explained by the specially chosen dataset, the chosen encodings that separability of datasets, as well as the robustness of the algorithm to noise in the quantum processor and the enough quality and stability of the quantum gates.
Additionally, since the classification accuracy is the same for the training and test datasets, we do not expect the overfitting problem in our case.

To further investigate how noise in considered encoding circuits with different circuit structures and their optimization modes affects the kernel matrix estimation procedure, we analyze the calculated average distances between experimentally obtained kernel matrices and ones calculated using a noiseless emulator (values in brackets in Tables~\ref{tab:results} and~\ref{tab:results-graphs}).
In the ideal case, when there is no noise in a real quantum processor, the distances should be equal to zero. 
From Table~\ref{tab:results} one can see that the distance grows with the number of two-particle gates in the circuits, in both optimized and non-optimized transpilation modes. For example, for non-optimized transpilation the smallest mean distance of 0.07 is achieved for the $R_Y$ encoding (no two-qubit gates), increasing to 0.33 for $R_Y+{\sf CX}$ encoding (4 two-qubit gates) and maximizing at amplitude encoding with the mean distance of 0.52 (8 two-qubit gates).
Since two-particle gates contribute the most to the total error, this is an expected behavior. 

On the other hand, optimized transpilation also significantly reduces the error and the distance between kernel matrices. 
A clear example is optimized transpilation for $R_Y+{\sf CX}$ encoding, where all two-particle gates are eliminated and distance is significantly decreased compared to non-optimized transpilation mode.
We note that for the amplitude encoding, where the number of two-particle gates cannot be reduced as much as in $R_Y+{\sf CX} $ encoding, the distance between kernel matrices remains almost the same for both transpilation modes.

\BLUE{
The main feature of the encoding method used for the second task, namely the classification of weighted graphs with $n$ vertices, is that it utilizes the feature of quantum entanglement and allows for the embedding of an Ising Hamiltonian spectrum of exponentially large $2^n$ size, within $2^n$ phases of components of a superposition state of $n$ qubits.
As shown in Table~\ref{tab:results-graphs}, the application of optimized transpiling significantly enhances the quality of kernel matrix reconstruction, but even without optimizations, the algorithm achieves $100\%$ accuracy on the datasets under consideration. 
This suggests that quantum devices may be a promising avenue for addressing this type of graph classification problems in the context of studying phase transitions, network analysis, and related fields.
We also note that the QAOA-inspired graph encoding~\eqref{eq:graph-encodings} employed can be extended by adding extra layers (alternating ``mixing'' and ``target'' ones), which can increase the expressiveness of the considered type of graph encoding.
}

The increase in the number of qubits employed allows for an expansion of the Hilbert space used for embedding classical data points. 
This, in turn, facilitates the loading of classical data and enables the processing of input classical data points with a higher dimensionality. Furthermore, leveraging the entanglement within a larger quantum register is anticipated to facilitate the solution of classification tasks, particularly in cases where the initial dataset structure does not permit a linear separation of classes.
Of course, the quality of implementing qubit gates is crucial, as otherwise we may not obtain reliable results.
We also note that selecting an effective quantum kernel function for a specific dataset remains a non-trivial challenge, similar to what is often encountered when employing classical SVM algorithms.

\section{Conclusion}

We performed a proof-of-principle quantum machine learning experiment using the developed trapped-ion-based quantum processor.
We considered classification problems for two datasets.
For the binary digit classification task, we considered three different quantum encodings of the dataset and compared results for both optimized and non-optimized transpilation modes. 
The first $R_Y$ encoding is the simplest encoding and has no two-qubit gates.
The second $R_Y+{\sf CX}$ encoding follows the same concept and has the same kernel matrix in the noiseless case, but additional ${\sf CX}$ gate introduces more noise within experimental results. 
As the third encoding we chose amplitude encoding, which embeds classical data to qubits' space. This encoding used entanglement of quantum states and employs the largest number of two-qubit gates.
A comparison between the three encodings with different noise levels for solving a single classification task with known results allowed one to determine the stability of the algorithm execution on real hardware.

\BLUE{
We also considered the problem of classifying weighted graphs based on the ground states of their corresponding Ising Hamiltonians, the reconstruction of which is known to be an NP-hard problem.
For this purpose, we applied a QAOA-inspired encoding scheme that uses $n$ of entangling gates for embedding the Ising spectrum of $2^n$ size into the probability amplitudes of an entangled state, exploiting the full dimensionality of $n$-qubit Hilbert space.
The encoding structure allows for consideration of non-optimized and optimized versions of the corresponding quantum-enhanced support vector machine circuits, consisting of $2n$ and $n$ entangling gates, respectively.}

In our experiments we showed that, 
despite the error in the kernel matrices, the estimation procedure increases with the raising of the two-qubit gates number in the circuits, the final result accuracy (classification accuracy) remains stable.
For each problem, we developed the classifier working with $100\%$ accuracy on both training and test sets.
Our results indicate the ability of the quantum processor to correctly solve basic, small-scale classification tasks considered.
As we expect, with the increase in capabilities, quantum processors can be utilized for solving machine learning tasks.

\section*{Acknowledgements}
This work was supported by the Russian Roadmap on Quantum Computing (Contract No. 868-1.3-15/15-2021, October 5, 2021).

\section*{Data availability}

The data that support the findings of this study are available from the corresponding author upon reasonable request.

\appendix

\section{Datasets} \label{app:data}

\begin{figure}
    \includegraphics[width=1.0\linewidth]{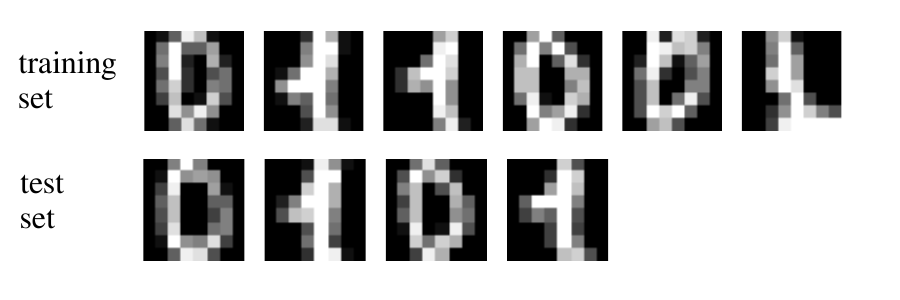}
    \caption{Training and test datasets of digits images. All images are naturally divided into two classes, depicting zeroes and ones, respectively.}
\label{fig:datasets}
\end{figure}

\begin{figure*}
    \centering
    \includegraphics[width=0.95\linewidth]{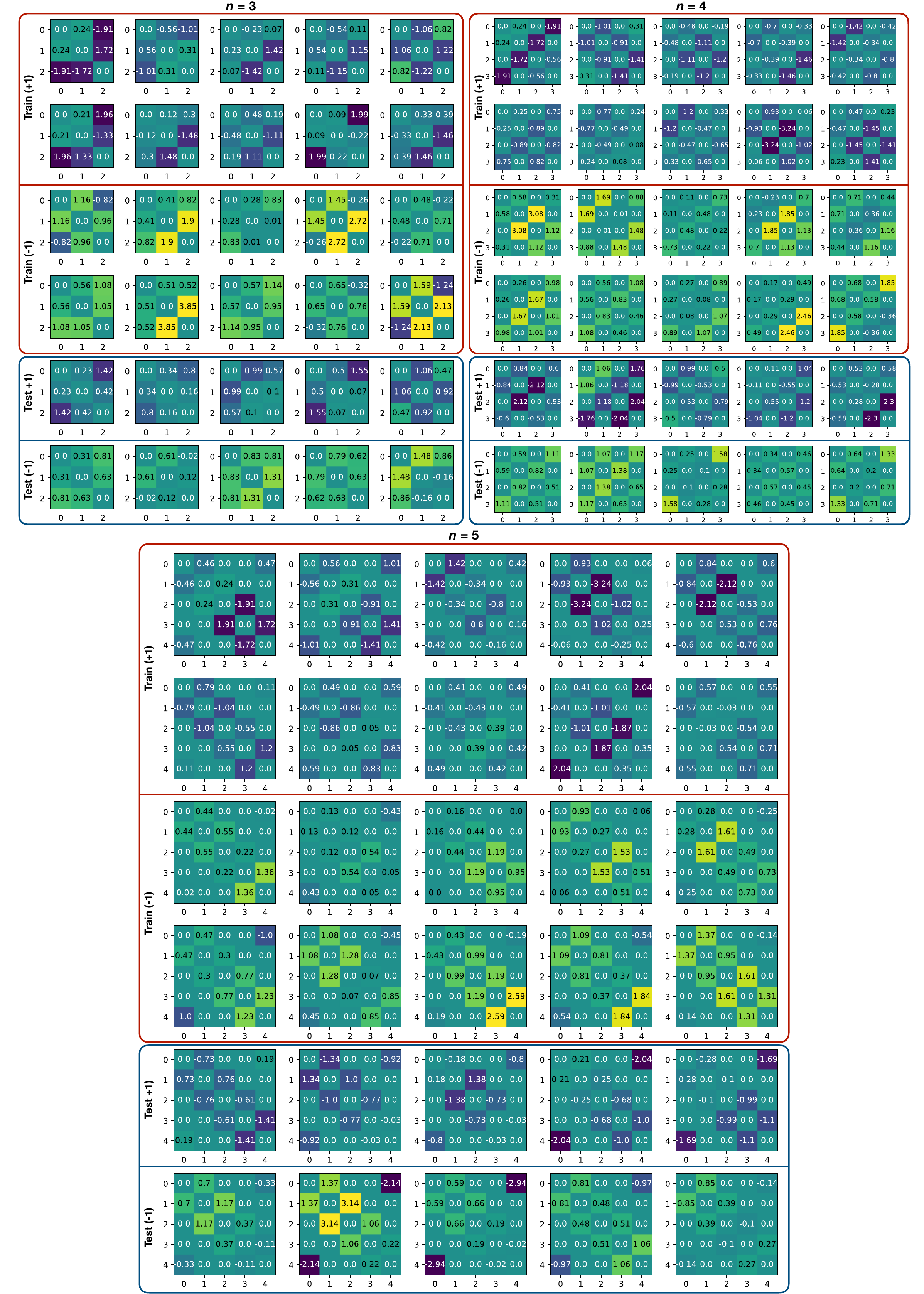}
    \caption{Weight matrices of training and test graphs used in the considered graph classification problems.}
    \label{fig:graphs_datasets}
\end{figure*}

In Fig.~\ref{fig:datasets} and Fig.~\ref{fig:graphs_datasets}, we present all the datasets used for the digit classification and graph classification problems, respectively.

\bibliography{bibliography.bib}

\end{document}